\documentclass[11pt,letterpaper,onecolumn]{article}
\usepackage[T1]{fontenc}
\usepackage{iftex}
\ifPDFTeX\usepackage[utf8]{inputenc}\fi
\usepackage{lmodern}
\usepackage[margin=1in]{geometry}
\usepackage{amsmath,amssymb}
\usepackage{graphicx}
\usepackage{booktabs,array,tabularx}
\usepackage[font=small,labelfont=bf]{caption}
\usepackage{placeins}
\usepackage{microtype}
\usepackage{xurl}
\usepackage[hidelinks]{hyperref}
\hypersetup{
  pdftitle={Transfer Learning and Experience Replay for Automotive Aerodynamic Surrogates},
  pdfauthor={Riddhiman Raut, Mehdi Bor, Katie Guo, Saakaar Bhatnagar, and Yin Yu},
  pdfsubject={Transfer learning and experience replay for automotive aerodynamic neural surrogates},
  pdfkeywords={Automotive aerodynamics, CFD, LSMART, transfer learning, continual learning, experience replay}
}
\title{Transfer Learning and Experience Replay for Automotive Aerodynamic Surrogates}
\author{\normalsize Riddhiman Raut, Mehdi Bor, Katie Guo,\\
        \normalsize Saakaar Bhatnagar, and Yin Yu\\[5pt]
        \normalsize Luminary, San Mateo, CA, USA}
\date{}

\begin{document}
\maketitle
\begin{abstract}
\noindent
Neural surrogates enable rapid prediction of automotive aerodynamic fields, but achieving high accuracy often demands extensive computational fluid dynamics datasets. Generating these datasets for each new vehicle program or expansion of the design space can require prolonged simulation campaigns. Reusing knowledge from existing simulations offers a way to limit this burden, provided that models adapt to unfamiliar geometries while maintaining predictive accuracy. This study evaluated transfer learning and experience replay in two independent automotive aerodynamics experiments using a mesh-free neural surrogate. First, models pretrained on 1,981 sport utility vehicle cases from SHIFT-SUV were fine-tuned on subsets of the SHIFT-Truck pickup dataset. Target training-set size was varied from 100 to 613 cases across the surface-only and joint surface and volume configurations to assess data efficiency. Fine-tuning reduced every reported field mean absolute error (MAE) relative to training from scratch at the same target budget. For joint surface and volume prediction, fine-tuning on 100 truck cases matched or improved on all four field MAEs obtained by training from scratch on 200 cases. Second, an independently trained model learned a base task using 332 DrivAerML cases and was adapted to a novel task using 31 cases with a previously unseen combination of design features. Experience replay incorporated cases from the base-task training set during adaptation. Full replay lowered field errors on 45 held-out base-task designs by approximately 36\% to 42\% relative to adaptation without replay; retaining half the base-task training set yielded reductions of approximately 30\% to 35\%. Together, these findings support transfer learning and experience replay as complementary strategies for data-efficient neural surrogate development across successive engineering tasks.
\end{abstract}


\section{Introduction}\label{sec:1}

Aerodynamic efficiency is a central objective in automotive design because it directly influences fuel consumption in gasoline vehicles and driving range in electric vehicles. Even modest reductions in drag can offer practical benefits: Mercedes-Benz estimates that decreasing the drag coefficient by 0.01, equivalent to ten drag counts, can increase long-distance electric vehicle range by approximately 2.5\% \cite{mercedes2025}. Identifying such improvements requires repeated evaluation of candidate geometries, creating a need for rapid aerodynamic assessment during vehicle development. In recent years, neural surrogate models have shown promise for rapid prediction of automotive aerodynamic fields, allowing more candidate designs to be assessed \cite{ananthan2024,li2023gino,wu2024transolver}. Public computational fluid dynamics (CFD) datasets have also expanded the resources available for training and benchmarking these models \cite{ashton2025drivaerml,elrefaie2024,ashton2024ahmedml,ashton2024windsorml}. Their accuracy, however, depends on data that represent the geometries and flow conditions relevant to the intended application. Producing these CFD datasets entails geometry preparation, mesh generation, numerical solution, and quality control across many designs. The resulting simulation campaigns can demand substantial time and computational resources, creating a barrier to surrogate deployment even when inference is fast. Reducing the amount of new simulation data required to achieve useful predictive accuracy is therefore a key challenge in making neural surrogates practical for iterative automotive design.

The problem of generating training data can worsen as vehicle programs evolve and introduce geometries beyond those represented in an existing surrogate's training set. An extensive CFD archive for one vehicle family may provide limited coverage of another, or of new geometric variations within the same family. Changes to the roofline, front fascia, or side mirrors, for example, can alter the surrounding flow and challenge the predictive accuracy of a model trained on earlier designs. Generating a large new dataset for each application would impose repeated demands on simulation resources. Reusing aerodynamic relationships learned from earlier simulations could reduce this additional data requirement, making the ability to adapt existing models to unfamiliar geometries an important step toward more data-efficient surrogate development.

To this end, transfer learning provides a general framework for applying previously acquired knowledge to a related learning problem \cite{pan2010}. For neural surrogate models, one implementation is full fine-tuning, in which pretrained weights initialize the target model and all parameters are updated using target data. Keum and Warey examined transfer across vehicle families using 20 target-family cases, evaluating surface fields and forces and finding that low-rank adaptation outperformed full fine-tuning in their setting \cite{keum2026}. These findings motivate examining full fine-tuning across target-data budgets, including joint surface and volume prediction and whether improved surface fields also yield lower drag error. However, fine-tuning on new data can reduce accuracy on previously learned designs, a phenomenon known as forgetting \cite{kirkpatrick2017,rolnick2019}. Maintaining earlier predictive capability therefore becomes an additional requirement when the model must serve an expanding design space.

Continual learning addresses this requirement by seeking to incorporate new information while preserving previously acquired knowledge \cite{kirkpatrick2017,rolnick2019}. Experience replay is one strategy: selected examples from earlier training data are retained and combined with new data during subsequent model updates \cite{rolnick2019}. Samuel and Ahmed benchmarked continual learning for engineering regression, including automotive aerodynamics \cite{samuel2025}. However, their automotive benchmarks focused on drag-coefficient prediction and did not assess spatially distributed surface or volume fields. Further investigation is therefore needed to determine how replay-memory size and case selection affect the accuracy of surface and volume field predictions on earlier designs after adaptation to new geometries.

To address these questions, this study used LSMART, a mesh-free neural surrogate model \cite{gao2026lsmart}, in two independent automotive aerodynamics experiments. In the first, models pretrained on sport utility vehicles (SUVs) were fine-tuned on pickups and compared with training from scratch across target-data budgets, using SHIFT-SUV and SHIFT-Truck \cite{shiftsuv2025,raut2026truck}. Both surface-only and joint surface and volume predictions were evaluated. The surface-only evaluation also assessed whether improved field predictions translated into lower drag-coefficient error. The second experiment adapted an independently trained model from a base task to a novel task within DrivAerML and evaluated field accuracy on held-out base-task designs across replay-memory sizes and case-selection strategies \cite{ashton2025drivaerml}. Together, the experiments examine how pretrained models and retained CFD cases can extend the value of existing simulation data as vehicle designs evolve.

Section~\ref{sec:2} describes the methodology. Section~\ref{sec:3} presents the transfer learning and experience replay results. Section~\ref{sec:4} discusses their implications for aerodynamic surrogate development. Section~\ref{sec:5} summarizes the findings and identifies directions for future work.

\section{Methodology}\label{sec:2}

\subsection{Neural Surrogate Model}\label{sec:2.1}

LSMART is a mesh-free neural surrogate model \cite{gao2026lsmart}, based on the SMART encoder--decoder architecture \cite{hagnberger2026smart}. The encoder builds a compact representation of the vehicle geometry from a surface point cloud that includes normals and signed-distance information. The encoder is initialized from a coarse subset of the surface points. At each layer, cross-attention gathers information from the input point cloud, and a multilayer perceptron refines the resulting features. Each decoder layer draws on the geometry features from the corresponding encoder layer. The final decoder layer maps those features to field values at the requested locations.

Two configurations were used in the transfer experiment. The surface-only model predicts pressure and wall shear stress over the vehicle surface. The joint surface and volume model also predicts pressure and velocity in the surrounding flow. Each configuration contains approximately 12.5 million trainable parameters. The continual-learning experiment used the joint configuration.

\subsection{Datasets and Data Preparation}\label{sec:2.2}

Both model configurations were pretrained on 1,981 full-scale SHIFT-SUV cases. The pickup training and evaluation sets were drawn from a pool of 712 SHIFT-Truck cases, including pickups with open and covered cargo beds. The SUV and truck cases were generated using delayed detached-eddy simulation (DDES) at fixed freestream speeds of 108 and 140 km/h, respectively. Vehicle geometries and CFD setups are detailed in the SHIFT publications \cite{raut2026truck,emory2026shift}.

The surface-only experiment used an approximately 90\%/10\% training/test allocation, with 613 training cases at the largest budget. The smaller datasets of 100, 200, 300, and 434 cases were subsets of this training pool, and the test set was held fixed across budgets. The joint surface and volume experiment used an approximately 80\%/10\%/10\% training/validation/test allocation, with 550 training cases at the largest budget. Its 100-, 200-, and 300-case datasets were subsets of the training pool, while the validation and test sets remained fixed. At each budget, fine-tuning and training from scratch used identical training cases and evaluation partitions. Varying training-set size while holding evaluation cases fixed tests data efficiency without confounding the comparison with changes in the evaluated geometries.

The continual-learning experiment used 484 DrivAerML cases \cite{ashton2025drivaerml}. In the dataset's parameter space, the novel task comprised 41 cases that simultaneously satisfied three conditions: a backlight angle of at least 69.1, a rear diffuser angle of at least 7.6, and a vehicle ride height of at most \(-6.3\). The remaining 443 cases formed the base task. Each condition individually, and each pair of conditions, was satisfied by cases in the full base-task partition, but no base-task case satisfied all three conditions together. The split therefore tested adaptation to a new combination of design features within a single passenger-car family.

The base task contained 332 training, 66 validation, and 45 held-out test cases; the novel task contained 31 training, 6 validation, and 4 held-out test cases (Table~\ref{tab:1}). These assignments were fixed throughout the replay comparisons. Validation and test cases were excluded from replay memory.

\begin{table}[tbp]
\centering
\caption{DrivAerML partition for the replay experiment. Validation cases select checkpoints; the four novel test cases are distinct from that selection population.}\label{tab:1}
\small
\renewcommand{\arraystretch}{1.18}
\begin{tabular}{lrrrr}
\toprule
\textbf{Task} & \textbf{Train} & \textbf{Validation} & \textbf{Test} & \textbf{Total} \\
\midrule
Base & 332 & 66 & 45 & 443 \\
Novel & 31 & 6 & 4 & 41 \\
Total & 363 & 72 & 49 & 484 \\
\bottomrule
\end{tabular}
\end{table}

Each CFD case was represented by randomly sampled points during training and evaluation. Truck cases used approximately 16,000 surface points and, for the joint configuration, 80,000 volume points. DrivAerML cases used approximately 16,000 surface and 32,000 volume points.

For SUV-to-truck transfer, input and output normalization statistics computed on the source data were retained during fine-tuning and inference.

\subsection{Transfer-Learning Procedure}\label{sec:2.3}

\begin{figure}[tbp]
\centering
\includegraphics[width=\linewidth]{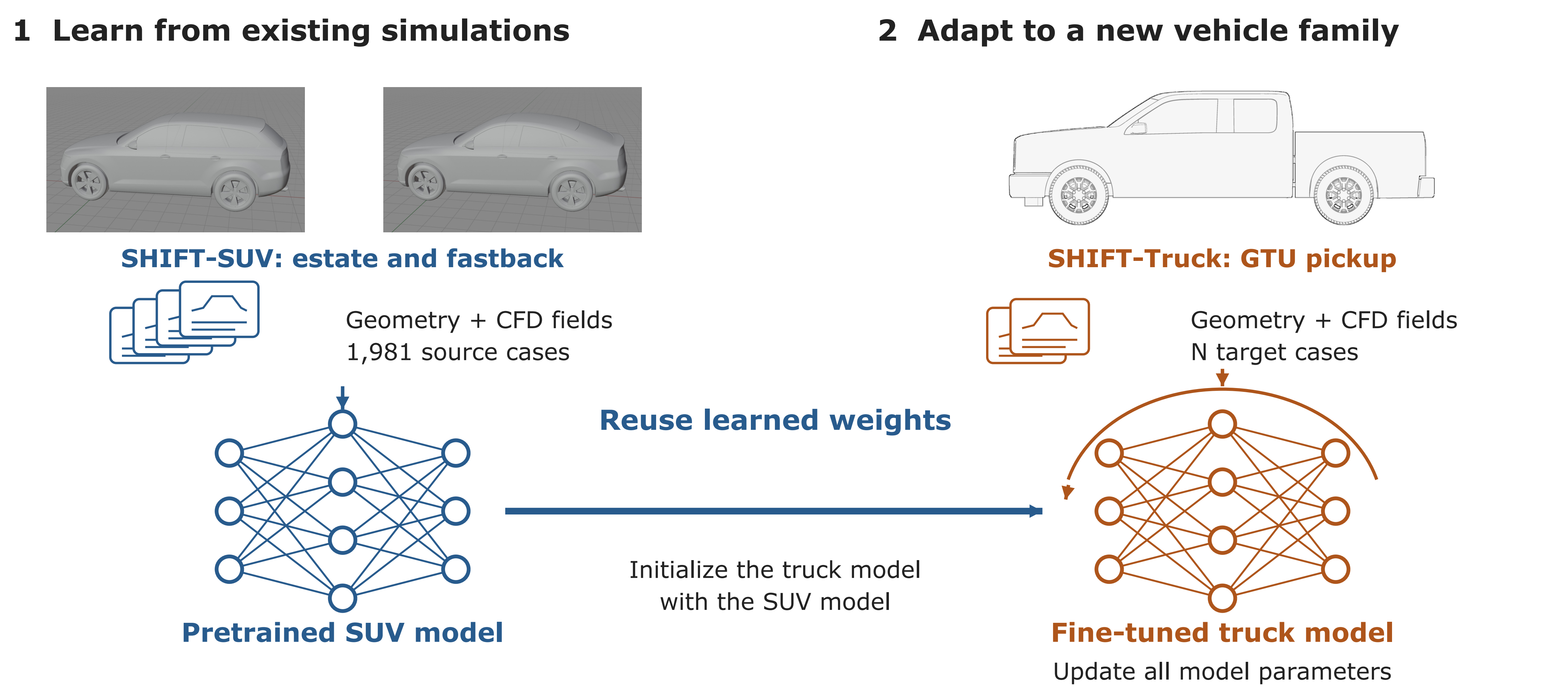}
\caption{The truck LSMART model is initialized with SUV-pretrained weights and then fine-tuned on truck CFD data. The baseline is trained from random initialization at the same target budget (Table~\ref{tab:2}). Model symbols are conceptual. SUV images are from Emory et al. \cite{emory2026shift}, Figure 1. The AeroSUV geometry is from the Research Institute for Automotive Engineering and Powertrain Systems Stuttgart (FKFS), and the images are from Luminary. The truck outline is adapted from Raut et al. \cite{raut2026truck}, Figure 1a, with the parameter annotations removed under \href{https://creativecommons.org/licenses/by/4.0/}{CC BY 4.0}. Vehicle views are shown at different scales.}\label{fig:1}
\end{figure}

Full fine-tuning and training from scratch were compared using the same truck training cases at each budget in Table~\ref{tab:2}. Separate SUV-pretrained checkpoints initialized the two configurations: the surface-only truck model was fine-tuned from a surface-only checkpoint, and the joint surface and volume truck model was fine-tuned from a combined checkpoint. Both checkpoints were pretrained on the same 1,981 SHIFT-SUV cases. All parameters were updated during fine-tuning. For each configuration, the baseline was trained from random initialization. Figure~\ref{fig:1} illustrates the transfer workflow.

\begin{table}[tbp]
\centering
\caption{Target training budgets and reported quantities in the two truck transfer experiments.}\label{tab:2}
\small
\renewcommand{\arraystretch}{1.18}
\begin{tabularx}{\linewidth}{@{}>{\raggedright\arraybackslash}p{0.20\linewidth} >{\raggedright\arraybackslash}p{0.27\linewidth} >{\raggedright\arraybackslash}X@{}}
\toprule
\textbf{Model} & \textbf{Training cases} & \textbf{Reported quantities} \\
\midrule
Surface only & 100, 200, 300, 434, 613 & Surface pressure, wall shear stress, and drag \\
Surface and volume & 100, 200, 300, 550 & Surface pressure and wall shear stress, plus volume pressure and velocity \\
\bottomrule
\end{tabularx}
\end{table}

SUV pretraining and truck training minimized a relative \(L_2\) field loss using the Lion optimizer \cite{chen2023lion}. The initial learning rate was \(1 \times 10^{-5}\) for SUV pretraining and truck training from scratch. For fine-tuning, it was \(5 \times 10^{-5}\). Table~\ref{tab:3} summarizes the training settings.

For each surface-only truck run, the checkpoint with the lowest training loss was retained. For the joint surface and volume runs, checkpoint selection was based on the lowest loss on a separate validation set. The selected models were evaluated on held-out test cases in both configurations.

\begin{table}[tbp]
\centering
\caption{Training settings for transfer and continual learning. Sampling counts are approximate.}\label{tab:3}
\small
\renewcommand{\arraystretch}{1.18}
\begin{tabularx}{\linewidth}{@{}>{\raggedright\arraybackslash}p{0.22\linewidth} >{\raggedright\arraybackslash}X@{}}
\toprule
\textbf{Setting} & \textbf{Value} \\
\midrule
Model size & Approximately 12.5 million trainable parameters \\
Optimizer & Lion with beta values of 0.9 and 0.99 and zero weight decay \\
Initial learning rate & \(1 \times 10^{-5}\) for SUV pretraining, training from scratch, and replay Stage 1. \(5 \times 10^{-5}\) for fine-tuning and replay Stage 2. \\
Schedule & 500 epochs with a cosine learning rate schedule updated each epoch and ending at zero. Warmup and early stopping disabled. \\
Batch and precision & One case per batch, fp32 \\
Gradient clipping & Norm limit of 10 and value limit of 10 \\
Truck sampling & 16,000 surface points and 80,000 volume points for the joint model \\
DrivAerML sampling & 16,000 surface points and 32,000 volume points \\
Training loss & Relative \(L_2\) for SUV pretraining and truck training. MAE for DrivAerML. \\
Checkpoint selection & Lowest validation loss for SUV pretraining and joint truck runs. Lowest training loss for surface-only truck runs. Lowest validation loss for the corresponding replay stage. \\
Truck training hardware & Eight GPUs \\
\bottomrule
\end{tabularx}
\end{table}

\subsection{Continual-Learning Procedure}\label{sec:2.4}

In Stage 1, LSMART was trained from random initialization on the 332 base-task training cases at an initial learning rate of \(1 \times 10^{-5}\). The checkpoint with the lowest loss on the 66 base-task validation cases was used to initialize every Stage 2 run.

\begin{figure}[tbp]
\centering
\includegraphics[width=\linewidth]{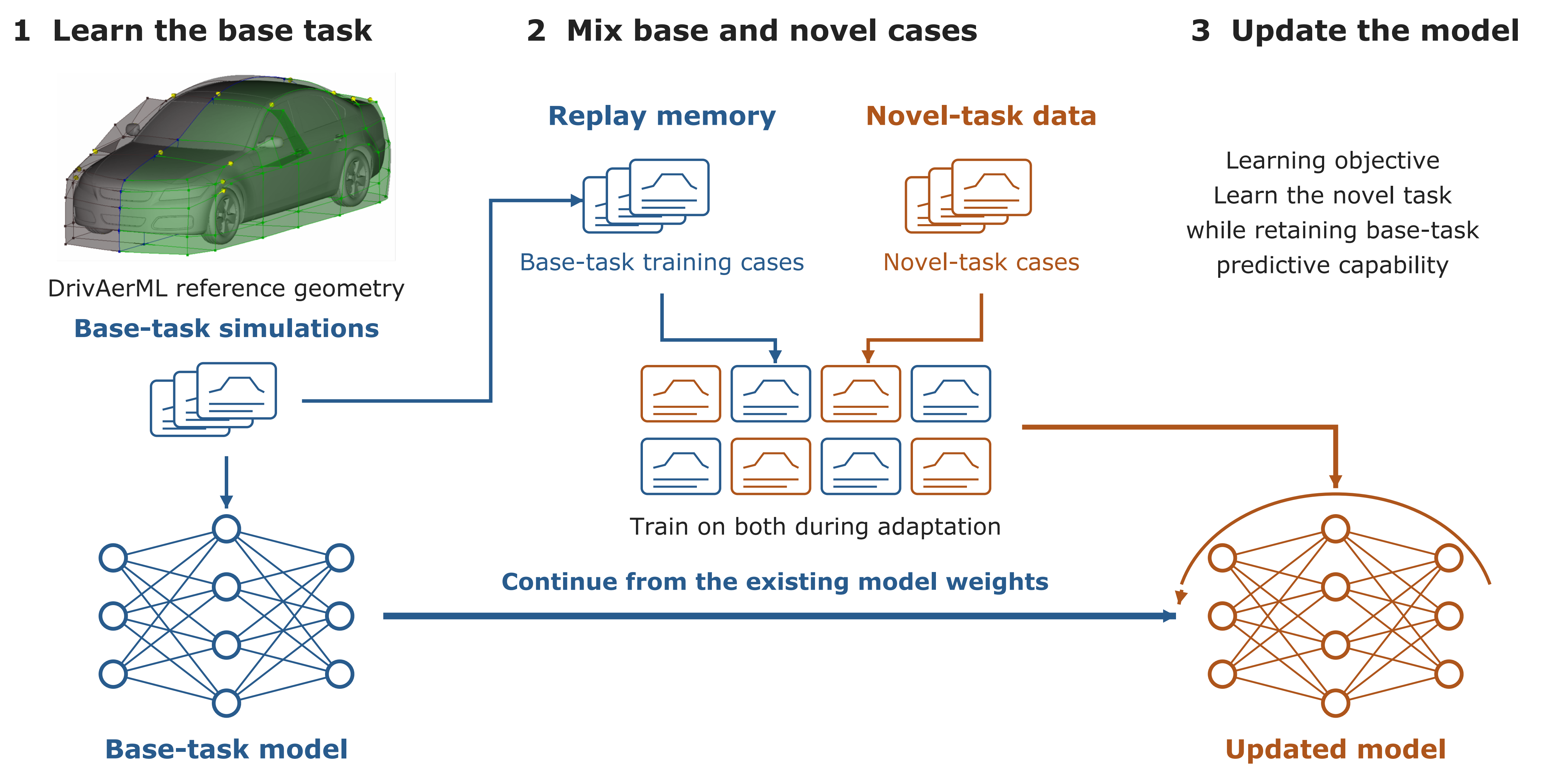}
\caption{Two-stage training procedure for continual adaptation. LSMART was first trained on base-task designs. The resulting checkpoint was fine-tuned on novel cases, with replay drawing retained cases from the base-task training set. Model symbols and card counts are conceptual. The representative DrivAerML geometry is from Ashton et al. \cite{ashton2025drivaerml}, Figure 2a, under \href{https://creativecommons.org/licenses/by/4.0/}{CC BY 4.0}.}\label{fig:2}
\end{figure}

In Stage 2, the same 31 novel training cases were used in every run. Base-task training cases were added at the replay memory sizes in Table~\ref{tab:4}, and the no-replay baseline used the novel cases alone. All model parameters were updated with an initial learning rate of \(5 \times 10^{-5}\). Both stages used MAE loss and the training settings in Table~\ref{tab:3}. Two NVIDIA H100 GPUs were used for every run in both stages. Figure~\ref{fig:2} illustrates the procedure.

For random replay, the base-task training cases were drawn uniformly without replacement to form a single ordering. For latent-diverse replay, an ordering was constructed by farthest-point sampling \cite{qi2017pointnet} using 768-dimensional geometry embeddings from a pretrained Point-MAE encoder \cite{pang2022pointmae}. The embeddings were computed once and held fixed. At each step, the case with the largest Euclidean distance to its nearest selected case was added. Each memory was formed from the start of its ordering, so larger memories retained the cases selected at smaller sizes. At full replay, both strategies used all 332 base-task training cases and shared a single result.

Every Stage 2 run followed the 500-epoch schedule, and the checkpoint with the lowest validation loss was retained. Each validation set contained the same six novel cases and the number of base-task cases specified in Table~\ref{tab:4}. Performance on base-task designs was evaluated on the same 45 held-out test cases for every configuration.

\begin{table}[tbp]
\centering
\caption{Stage 2 dataset composition and scheduled case presentations. Replay memory is expressed as a percentage of the 332 base-task training cases. Each setting includes 31 novel training and six novel validation cases. Case presentations are calculated as the total number of training cases multiplied by 500 epochs.}\label{tab:4}
\small
\renewcommand{\arraystretch}{1.18}
\begin{tabularx}{\linewidth}{@{}*{5}{>{\centering\arraybackslash}X}@{}}
\toprule
\textbf{Replay memory (\%)} & \textbf{Base training cases} & \textbf{Total training cases} & \textbf{Case presentations} & \textbf{Base / total validation cases} \\
\midrule
0 & 0 & 31 & 15,500 & 0 / 6 \\
10 & 33 & 64 & 32,000 & 7 / 13 \\
25 & 83 & 114 & 57,000 & 17 / 23 \\
50 & 166 & 197 & 98,500 & 33 / 39 \\
75 & 249 & 280 & 140,000 & 50 / 56 \\
100 & 332 & 363 & 181,500 & 66 / 72 \\
\bottomrule
\end{tabularx}
\end{table}

\begin{samepage}
\subsection{Evaluation Metrics}\label{sec:2.5}

Agreement with the reference CFD solution is quantified using mean absolute error (MAE) as defined in Eq.~\eqref{eq:1}. For each output channel \(q\), absolute errors are averaged over the sampled points within each case and then across the \(N\) evaluation cases, giving equal weight to each vehicle case.

\nopagebreak[4]
\begin{equation}\label{eq:1}
\operatorname{MAE}_{q} = \frac{1}{N}\sum_{i=1}^{N}\frac{1}{n_i}\sum_{j=1}^{n_i}\left|\widehat{q}_{ij}-q_{ij}\right|.
\end{equation}
\end{samepage}

Here, \(q_{ij}\) and \({\widehat{q}}_{ij}\) are the reference and predicted values at point \(j\) in case \(i\), and \(n_{i}\) is the number of sampled points in that case. The reported wall shear stress MAE is the mean of the \emph{x}, \emph{y}, and \emph{z} component MAEs. Pressure and wall shear stress errors are reported in pascals, and velocity errors are reported in meters per second.

For the surface-only truck configuration, the predicted pressure and wall shear loads are integrated over the vehicle surface to obtain drag. The drag-coefficient MAE is calculated using Eq.~\eqref{eq:2}

\nopagebreak[4]
\begin{equation}\label{eq:2}
\operatorname{MAE}_{C_D} = \frac{1}{N}\sum_{i=1}^{N}\left|\widehat{C}_{D,i}-C_{D,i}\right|,
\end{equation}

where \({\widehat{C}}_{D,i}\) and \(C_{D,i}\) are the predicted and reference drag coefficients for case \(i\). One plotted count in Figure~\ref{fig:3} corresponds to \(10^{-3}\) in \(C_D\).

The relative change in MAE with respect to the reference model is calculated using Eq.~\eqref{eq:3}

\nopagebreak[4]
\begin{equation}\label{eq:3}
\Delta E\;(\%) = 100\times\frac{E_{\mathrm{method}}-E_{\mathrm{ref}}}{E_{\mathrm{ref}}},
\end{equation}

where \(E_{\mathrm{method}}\) and \(E_{\mathrm{ref}}\) denote the MAEs of the evaluated model and its reference. The reference model is trained from scratch for truck transfer and adapted without replay for continual learning. Negative values of \(\Delta E\) indicate a reduction in error. Comparisons across truck training budgets are made within each model configuration. All configurations in the replay comparison are evaluated on the same 45 held-out base-task designs.

\FloatBarrier

\section{Results}\label{sec:3}

\subsection{Surface-Only Transfer and Drag Prediction}\label{sec:3.1}

\begin{figure}[tbp]
\centering
\includegraphics[width=\linewidth]{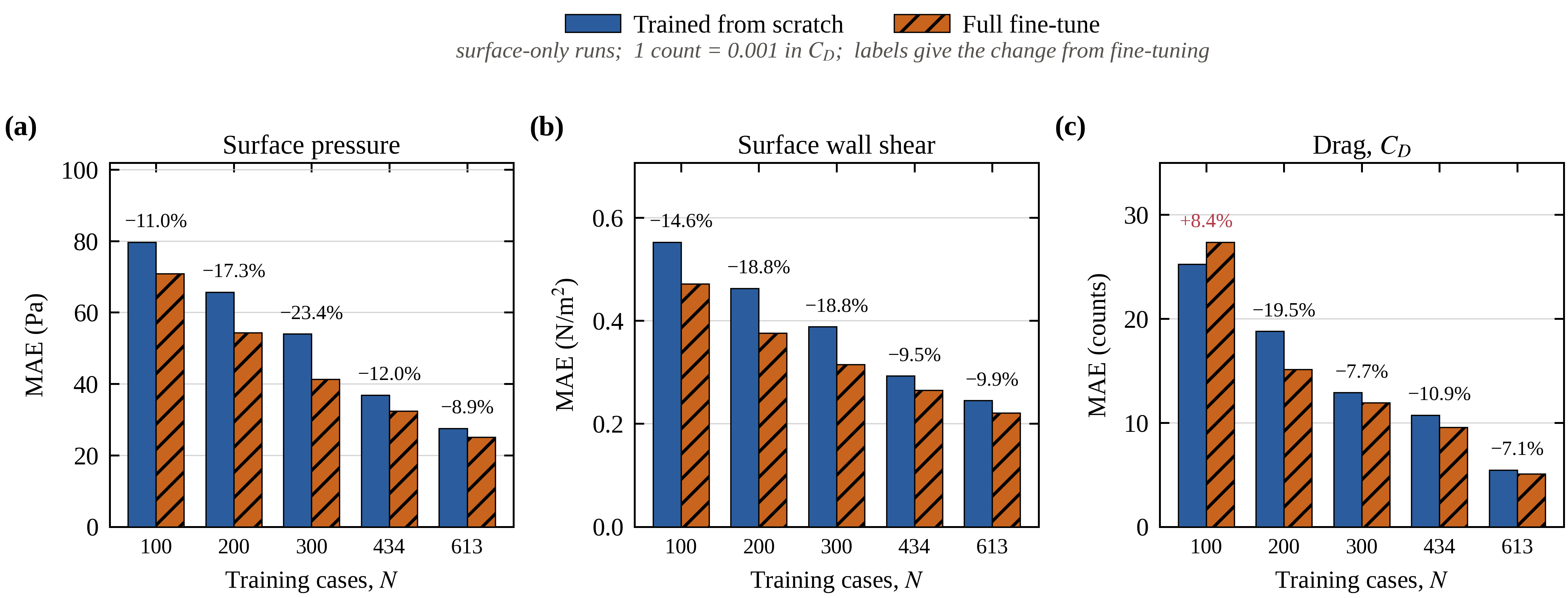}
\caption{Surface pressure, wall shear stress, and drag-coefficient MAEs for the surface-only truck model. Percent changes are relative to training from scratch at the same budget. One plotted drag count corresponds to \(10^{-3}\) in \(C_D\).}\label{fig:3}
\end{figure}

Fine-tuning the SUV-pretrained model improved both surface fields at every tested truck training budget (Figure~\ref{fig:3}). Relative to training from scratch, pressure MAE fell by 8.9\% to 23.4\% and wall shear stress MAE fell by 9.5\% to 18.8\%. For both training procedures, absolute field errors decreased as more truck cases were added. The relative fine-tuning advantage was smaller at the largest budget than at the intermediate budgets, although it did not decrease monotonically. This pattern is consistent with reduced dependence on source-task information as more target data become available.

The improvement extended to drag prediction at the four larger budgets, where drag-coefficient MAE fell by 7.1\% to 19.5\%. At 200 cases, for example, the error decreased from approximately 0.0188 to 0.0151. The 100-case comparison was the exception, with drag-coefficient MAE increasing by 8.4\% despite lower errors in both surface fields. Lower average field errors do not necessarily yield lower drag error because drag integration depends on the spatial weighting and signed cancellation of pressure and shear contributions.

\subsection{Transfer Learning for Joint Surface and Volume Prediction}\label{sec:3.2}

\begin{figure}[tbp]
\centering
\includegraphics[width=\linewidth]{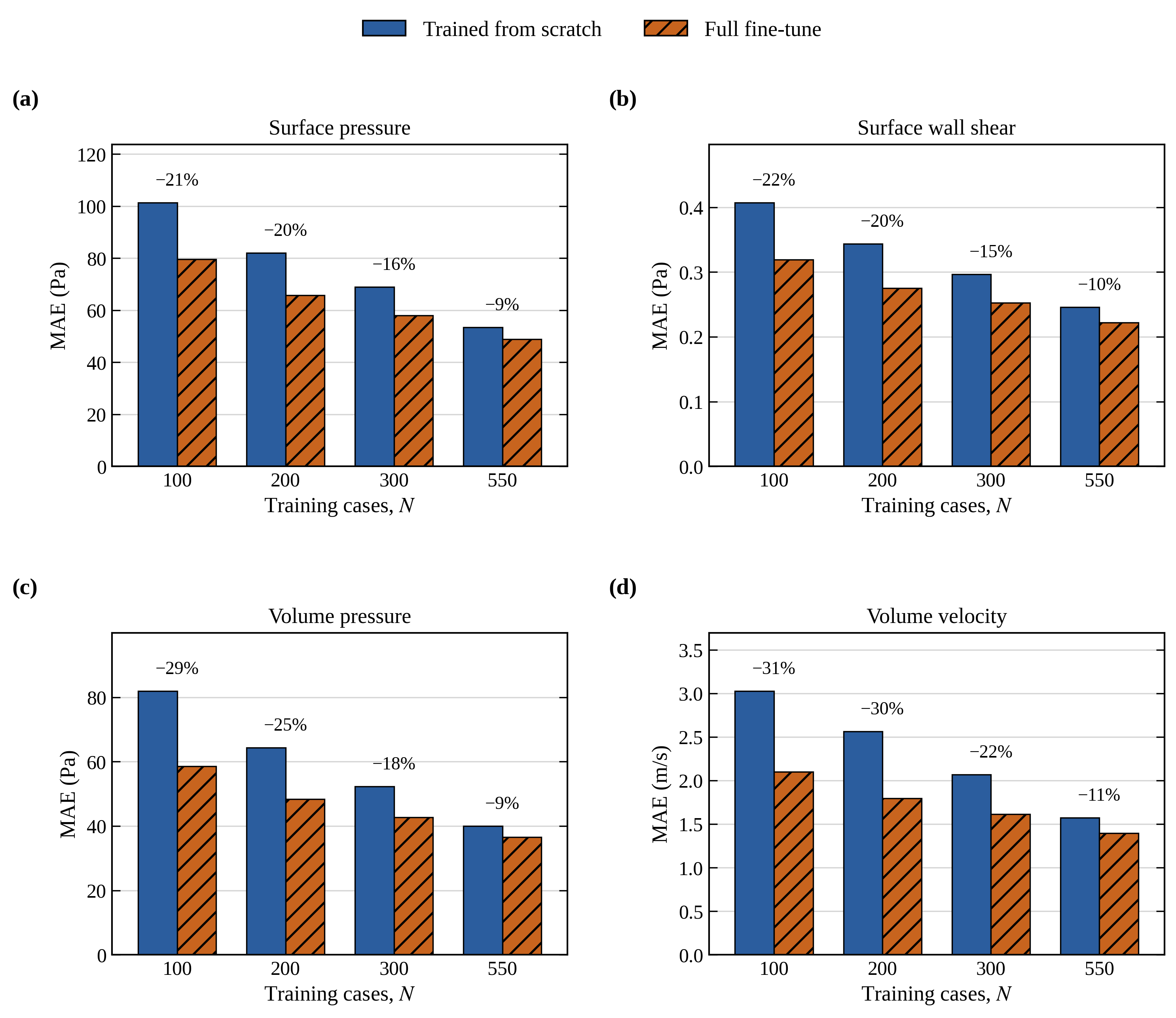}
\caption{Surface pressure (a), wall shear stress (b), volume pressure (c), and velocity (d) MAEs for the joint truck model. Percent changes are relative to training from scratch at the same budget.}\label{fig:4}
\end{figure}

When surface and volume fields were predicted jointly, the benefit of fine-tuning extended to the surrounding flow. All four field MAEs were lower than those of the corresponding models trained from scratch at every tested budget (Figure~\ref{fig:4}). The largest reductions occurred at 100 cases, where volume pressure and velocity MAEs fell by 29\% and 31\%, respectively. At 550 cases, reductions of 9\% to 11\% were observed across the four fields.

Comparisons across budgets show how an existing SUV checkpoint can reduce the demand for new truck simulations. Fine-tuning with 100 truck cases matched or improved on all four field MAEs obtained by training from scratch with 200 cases. Similarly, fine-tuning with 200 cases gave lower errors across all four fields than training from scratch with 300 cases. For all four fields, the relative advantage of fine-tuning narrowed as the target training set grew, consistent with a reduced contribution from pretraining when more target data were available.

\subsection{Continual Learning with Experience Replay}\label{sec:3.3}

\begin{figure}[tbp]
\centering
\includegraphics[width=\linewidth]{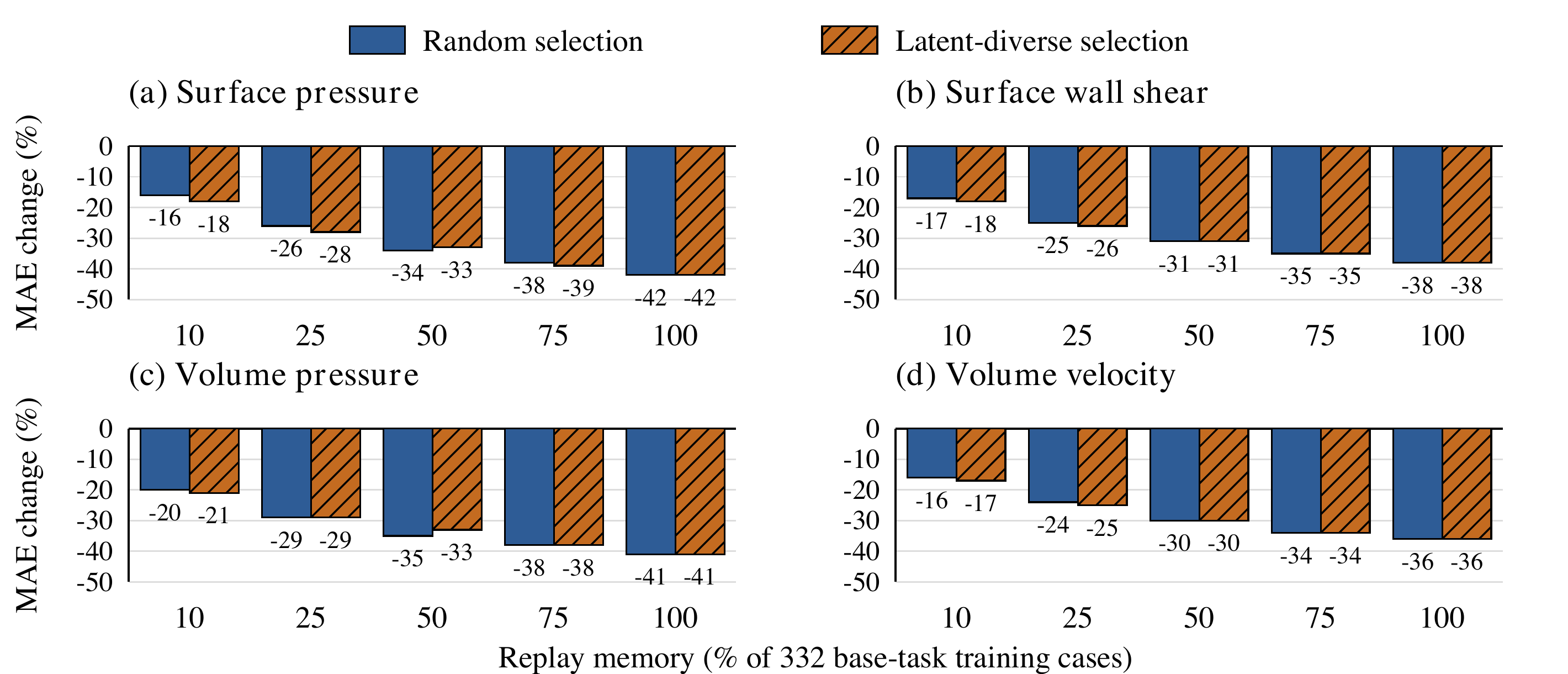}
\caption{Changes in field MAE on 45 held-out base-task DrivAerML designs after Stage 2 adaptation, relative to adaptation without replay. Random and latent-diverse replay use the same case counts at each memory fraction and share the full-memory result. Percent changes are rounded to whole numbers.}\label{fig:5}
\end{figure}

The DrivAerML experiment examined how replay affected predictions on the base task after adaptation to the novel task. At every tested memory size, replay reduced all four field MAEs on the 45 held-out base-task designs relative to adaptation without replay (Figure~\ref{fig:5}).

Retaining 10\% of the base-task training cases reduced errors by 16\% to 21\%. At half memory, retaining 166 base-task cases alongside the 31 novel cases brought reductions of 30\% to 35\%. Doubling the memory from half to full replay added a further 6 to 9 percentage points, giving reductions of 36\% to 42\% relative to adaptation without replay.

The two selection strategies yielded similar reductions at matched memory sizes, with a small advantage for latent-diverse selection at the two smallest memories. At 10\% memory, latent-diverse selection improved the reported error reductions by 1 to 2 percentage points across all four fields; at 25\%, it improved three fields by 1 to 2 percentage points and tied for volume pressure. This pattern is consistent with the intended benefit of covering more of the geometry embedding space when few cases can be retained. The advantage was not consistent at larger memories: random selection gave the larger reduction in volume pressure error at 50\%. These small differences describe the reported runs and do not establish a statistically reliable selector advantage. Across the tested settings, error reductions varied more with the amount of data retained than with the selection strategy.

\section{Discussion}\label{sec:4}

A surrogate used for early aerodynamic screening must reach useful accuracy while the database for a new vehicle program is still small. The aim is to distinguish the drag changes associated with candidate geometries so that designers can decide which changes to pursue. The pickup study shows how an existing checkpoint can reduce the amount of new data needed for adaptation. Fine-tuning the SUV-pretrained model on 100 truck cases matched or improved on all four field MAEs obtained by training from scratch on 200 cases. Fine-tuning also lowered drag-coefficient MAE at four of the five matched budgets in the surface-only experiment.

A different requirement arises when an updated surrogate must continue to serve earlier designs as well as new ones. Experience replay addresses this use case by bringing selected earlier simulations into training alongside the new cases. In DrivAerML, retaining half the base-task training set reduced field MAEs on held-out base-task designs by 30\% to 35\% relative to adaptation without replay. This result supports the use of replay when accuracy on earlier designs remains part of the modeling task.

\section{Conclusions}\label{sec:5}

This study evaluated transfer learning and continual learning as complementary approaches to training neural surrogates for automotive aerodynamics. The transfer learning experiment examined whether an SUV-pretrained LSMART model could reach comparable accuracy on pickup trucks with fewer simulations than training from scratch. The continual learning experiment addressed a different stage of model development, when an updated surrogate must continue to serve earlier designs. In that experiment, experience replay was evaluated while adapting a DrivAerML model from the base task to 31 novel-task cases. The main findings were as follows.

\begin{itemize}
\item
  \textbf{Fine-tuning improved field accuracy at every tested training budget.} Relative to training from scratch, fine-tuning lowered the mean absolute error (MAE) of every predicted field in both the surface-only and joint surface and volume configurations. With 100 pickup cases, the joint model's four field MAEs decreased by 21\% to 31\%.
\item
  \textbf{The joint model reached comparable field accuracy with fewer pickup cases.} Fine-tuning on 100 cases matched or improved all four field MAEs achieved by training from scratch on 200 cases. Fine-tuning on 200 cases also outperformed training from scratch on 300 cases across all four fields.
\item
  \textbf{Fine-tuning improved drag prediction at four of the five surface-only training budgets.} Drag-coefficient MAE decreased by 7.1\% to 19.5\% at the four larger budgets. At 100 cases, drag error increased despite lower surface field errors.
\item
  \textbf{Experience replay improved predictions on earlier designs after adaptation.} Relative to adaptation without replay, retaining half of the base-task training set reduced field MAEs by 30\% to 35\% on 45 held-out base-task designs. Full replay yielded reductions of 36\% to 42\%. Neither random nor latent-diverse selection consistently yielded lower errors.
\end{itemize}

Comparing predicted and CFD-derived drag differences between pairs of candidate geometries would test whether the surrogate captures the direction and magnitude of the changes that guide design selection. Freezing the encoder or decoder during fine-tuning would test whether updating fewer parameters can shorten training while maintaining accuracy. Comparing the effects of freezing the encoder versus the decoder would also help identify which parts of the learned mapping can be reused across vehicle families.

Taken together, the experiments show how prior simulations can support two stages of surrogate development. A pretrained model can reach comparable field accuracy with fewer simulations from a new vehicle family, while replay can improve predictions on earlier designs as the design space expands. By extending the value of costly simulations across successive modeling tasks, these approaches offer a route toward more data-efficient scientific machine learning and could make neural surrogates practical earlier in automotive aerodynamic development.

\FloatBarrier
\paragraph{Use of generative AI.} Generative AI tools were used to assist with grammar checking and text polishing to improve the clarity and readability of the manuscript. The authors accept full responsibility for the accuracy, integrity, and content of the final manuscript.


\end{document}